%% file: sigma2009c.tex
\documentclass[twoside]{article}
\usepackage{fleqn,espcrc2}
\usepackage{graphicx}
\usepackage{here}

\newcommand{\AmS}{{\protect\the\textfont2
    A\kern-.1667em\lower.5ex\hbox{M}\kern-.125emS}}										
\def\beq{\begin{equation}}
\def\eeq{\end{equation}}
\def\bea{\begin{eqnarray}}
\def\eea{\end{eqnarray}}
\def\bq{\begin{quote}}
\def\eq{\end{quote}}

\def\nnb{\nonumber}
\def\ga{\left(}
\def\dr{\right)}

\def\rar{\rightarrow}
\def\lrar{\Longrightarrow}
																
\def\nnb{\nonumber}
\def\la{\langle}
\def\ra{\rangle}
\def\nin{\noindent}
\def\ba{\vspace*{-0.2cm}\begin{array}}
\def\ea{\end{array}\vspace*{-0.2cm}}

\def\b{$\bullet~$}
\def\als{\alpha_s}

\def\gg2{ \la\alpha_s G^2 \ra}
\def\gg3{g^3f_{abc}\la G^aG^bG^c \ra}
\def\ggg4{\la\als^2G^4\ra}

\def\calD{ {\cal D} }
\def\ftilde{\tilde f}

\def\therho{\theta\rho}

\newcommand{\pipi}{\mbox{$\pi\pi$}}

\title
{\bf{\boldmath
{\Large Gluonium nature of the $\sigma/f_0(600)$  from its coupling  to $K\bar K$ } }}

\author{
R. Kami\'nski \thanks{{\it E-mail addresses:} Robert.Kaminski@ifj.edu.pl (R. Kaminski), gerard.mennessier@lpta.univ-montp2.fr (G. Mennessier), snarison@yahoo.fr (S. Narison).
} \address {\footnotesize Department of Theoretical Physics, H. Niewodnicza\'nski Institute of Nuclear Physics PAN, PL 31-342 Krak\'ow, Poland, },
G. Mennessier  $^{\rm{b}}$ and S. Narison \thanks{Corresponding author.}\address {\footnotesize Laboratoire
de Physique Th\'eorique et Astroparticules, CNRS-IN2P3 \& Universit\'e
de Montpellier II, Case 070, Place Eug\`ene
Bataillon, 34095 - Montpellier Cedex 05,
France.},
}

\begin{document}

\pagestyle{myheadings}
\markright{ }
\begin{abstract}
\noindent
We extract the $K^+K^-$ couplings of the isoscalar scalar mesons $\sigma/f_0(600)$ and $f_0(980)$
from $\pi\pi\to \pi\pi / K\bar K$ scatterings and found: $|g_{\sigma K^+K^-}|/ |g_{\sigma \pi^+\pi^-}|\simeq 0.8$ and  $|g_{f_0 K^+K^-}|/ |g_{f_0 \pi^+\pi^-}|\simeq 1.7$.
These results, together with the tiny  ``direct" $\gamma\gamma$ width of the $\sigma$ and its large hadronic width, are a strong indication for the  gluonium/glueball nature of the $\sigma$-meson,
 as predicted by QCD spectral sum rules (QSSR)  $\oplus$ some low-energy theorems (LET), while some other assignements ($\bar\pi\pi$ molecule, tetraquark state and ordinary $\bar qq$ meson) do not satisfy simultaneously these requirements from the data.  
 These properties suggest that the $\sigma$ can be  a scalar meson associated to the $U(1)_V$ conformal anomaly like is the $\eta'$-meson for the $U(1)_A$ anomaly. 
 \end{abstract}
\maketitle
\vspace*{-1.5cm}
\section{Introduction}
\vspace*{-0.25cm}
 \nin
Understanding the nature of scalar mesons in terms of quark and gluon
constituents is a long standing puzzle in QCD  \cite{MONTANET,SN09}. The problem here is that
 some 
states 
are very broad
($\sigma$ and $\kappa$(if confirmed) mesons) and others are close to an inelastic
 threshold
($f_0(980)$, $a_0(980)$), which makes their interpretation
 difficult. 
 One might expect that
the decay rate of these mesons into two photons could provide an
 important 
information
about their intrinsic composite structure. Indeed, a recent analysis of  $\gamma\gamma \to \pi\pi$ data \cite{MNO} indicates that
the $\sigma/f_0(600)$ (hereafter called $\sigma$) meson could be such a gluonic 
 resonance. In this paper, we pursue the test of its nature by studying
 its couplings (and that of the $f_0(980)$) to $K^+K^-$ and $\pi^+\pi^-$.  
 
\vspace*{-0.3cm}
\section{Gluonium nature of the $\sigma$ from $\pi\pi/\gamma\gamma\to\pi\pi$ }
\label{sec:qssr}
\vspace*{-0.25cm}
 \nin
The existence of glueballs/gluonia is a characteristic prediction of
 QCD and some
scenarios have been developed already back in 1975 \cite{MIN}. 
Today, there is agreement that such states exist in QCD and the
 lightest state has
quantum numbers $J^{PC}=0^{++}$. 
QSSR \cite{SNG0,VENEZIA,SNG,SN06,STEELE,SN09} determinations of its mass found$^1$ \footnotetext[1]{QSSR also requires the existence 
of a higher gluonium mass of about (1.5-1.6) GeV from a consistency of the subtracted and unsubstracted sum rules.}:
\beq
M_{\sigma_B}\simeq  1~{\rm GeV}~,
\eeq
confirmed recently by lattice
simulation using dynamical fermions \cite{MICHAEL} and a strong coupling
calculation \cite{FRASCA}. 
Some phenomenological implications of scalar glueball have been studied in the literature \cite{SNG0,VENEZIA,SNG,SN06,BN,SNHYB,DOSCHGLUE,MINK,CLOSE}.
The analysis of the $\gamma\gamma\to\pi\pi$ data \cite{MNO} using an improved model of \cite{mennessier} leads to the (model-dependent) ``partial" $\gamma\gamma$ widths:
\bea
\Gamma_{\sigma\to\gamma\gamma}^{\rm dir}&\simeq& (0.13\pm 0.05)~{\rm keV}~,\nnb\\
\Gamma_{\sigma\to\gamma\gamma}^{\rm resc}&\simeq& (2.7\pm 0.4)~{\rm keV}~,
 \label{radpart}
\eea
where $dir$ and $resc$ refer respectively to the direct coupling of the $\sigma$ resonance 
and to the rescattering term obtained using an unitarized Born amplitude. This leads to
the (model-independent) total $\gamma\gamma$ width (direct + rescattering):
 \beq
  \Gamma_{\sigma\to\gamma\gamma}^{\rm tot}\simeq (3.9\pm 0.6)~{\rm keV}~,
 \label{radtot}
\eeq
which is in agreeement with the results from the existing fits in
\cite{PENNINGTON,PENN1,OLLER,PRADES,ZHENG09}. The previous results in Eqs. (\ref{radpart}) and (\ref{radtot}) have been obtained at the complex pole obtained in \cite{MNO} and given in Table \ref{tab:sigma}, with the corresponding residue:
\beq
g_\pi \simeq  0.06 -{\rm i}~0.50~{\rm GeV}~.
\label{eq:pipiwidthfit}\label{eq:gpi}
\eeq
 \vspace*{-0.75cm}
{\scriptsize
\begin{table}[hbt]
\setlength{\tabcolsep}{0.2pc}
 \caption{\scriptsize    Mass and 1/2 width in MeV of the $\sigma$ meson in the complex plane. 
    }
\begin{tabular}{lll}
&\\
\hline
Processes&$M_\sigma -i\Gamma_\sigma/2$&Refs.   \\
\hline
$\pi\pi\to \pi\pi/K\bar K$&$422 -{\rm i}~290$&\cite{MNO}\\
&$441^{+16}_{-8} -{\rm i}~272^{+9}_{-15}$&\cite{leutwyler}\\
&$461 \pm 15 -{\rm i} ~(255 \pm 16)$&\cite{GKPY}\\
$J/\psi\to \omega\pi\pi$&$541\pm 39 -{\rm i}~(222\pm 42)$& \cite{BES2}\\
$D^+\to\pi^+\pi^-\pi^+$ &$478^{+24}_{-23} \pm 17-{\rm i}~(162^{+42}_{-40} \pm 21)$&\cite{E741}\\
\hline
\end{tabular}
\label{tab:sigma}
\end{table}
}
\vspace*{-0.5cm}
\nin
which is in agreement with the other results given there. These values of the $\sigma$ meson parameters in the complex plane, when 
appropriately translated into the real axis, become at the {\it on-shell mass} $M_\sigma^{\rm os}\approx 0.92$ GeV(see section \ref{sec:let}):
\beq
 \Gamma_{\sigma\to\pi\pi}^{\rm os}\approx 1.02~{\rm GeV}~,~
\Gamma_{\sigma\to \gamma\gamma}^{\rm os,dir}
\approx (1.0\pm 0.4)~{\rm keV}~,
\eeq
which are in a remarkable agreement with the QSSR  \cite{SVZ,SNB} and LET predictions obtained in the real axis $^2$\footnotetext[2]{These LET have been used earlier in \cite{NOVIKOV,CHANO}.} for a ``bare/unmixed"  gluonium/glueball $\sigma_B$ state having the mass \cite{SNG0}:
\beq
M_{\sigma_B}\simeq (0.95\sim 1.10) ~{\rm GeV}~,
\eeq
and the widths \cite{VENEZIA,SNG,SN06} from the couplings in Table \ref{tab:S2}:
\bea
\Gamma_{\sigma_B\to\pi^+\pi^-}&\simeq& 0.5~ {\rm GeV}~,\nnb\\
\Gamma_{\sigma_B\to\gamma\gamma}&\simeq& (0.2\sim 0.6)~{\rm keV}~.
\eea
 The large $\sigma$-width into $\pi\pi$ indicates a strong violation  of the OZI rule in this channel and 
 signals large non-perturbative effects in its treatment. This large hadronic width also disfavours its $\bar qq$ interpretation.
{\scriptsize
\begin{table}[hbt]
\setlength{\tabcolsep}{0.3pc}
 \caption{\scriptsize    QSSR ant LET predictions for the modulus of the hadronic couplings in GeV and $\gamma\gamma$ widths in keV of a scalar meson having a mass of 1 GeV for gluonium \cite{VENEZIA,SNG,SN06}, $\bar qq$ meson \cite{SNA0,SNG,SN06,SNMENES} and four-quark states \cite{SNA0}. 
    }
\begin{tabular}{lcccccc}
&\\
\hline
Meson&$g_{S\pi^+\pi^-}$ &$g_{ SK^+K^-}$&$\Gamma_{S\gamma\gamma}$   \\
\hline
$\sigma_B\equiv gg$&5&$g_{\sigma_B\pi^+\pi^-}$&$0.2\sim 0.6$\\
$S_2\equiv {1\over\sqrt{2}}(\bar uu+\bar dd)$&2.5&${1\over 2}g_{S_2\pi^+\pi^-}$&${25\over 9}\Gamma_{a_0\gamma\gamma}$ \\
$S_3\equiv \bar ss$&$g_{ S_3K^+K^-}$&$2.7\pm 0.5$&${1\over 9}\Gamma_{a_0\gamma\gamma}$\\
$(\bar qq) (q\bar q)$&&&$0.4\times 10^{-3}$\\
\hline
\end{tabular}
\label{tab:S2}
\end{table}
}
\nin
 In fact, using the hadronic couplings in Table \ref{tab:S2}, QSSR predicts for a $S_2\equiv 1/\sqrt{2}(\bar uu+\bar dd)$, with mass of 1 GeV:
 \beq
\Gamma_{S_2\to\pi^+\pi^-} \equiv  {|g_{S_2
  \pi^+\pi^-}|^2\over {16\pi M_{\sigma_B}}}\sqrt{ 1-{4m^2_\pi \over
 M^2_{S_2}}}\simeq 120~{\rm MeV}~. 
 \label{eq:scalarwidth}
\eeq
Using the ${a_0\to\gamma\gamma}$ width of about 2 keV from a QSSR analysis of a quark triangle loop vertex including the $\la \bar\psi\psi \ra$ condensate contribution \cite{SNA0}\footnote{A kaon tadpole loop mechanism leads to smaller value of the $a_0\to\gamma\gamma$ width of about 0.25 keV \cite{BN}.}, one predicts a $S_2\to\gamma\gamma$ width of about 5 keV (an analogous result has been obtained in \cite{ROSNER})\footnote{However, an approach based on the scalar anomaly leads to a smaller value of about 0.2 keV \cite{LANIK}.}and a  $S_3\to\gamma\gamma$ width of about 0.2 keV (see also \cite{BARNES}).  In the same way, QSSR also  predicts, for a four-quark state, having the same mass of 1 GeV \cite{LATORRE,SNA0}, a $\gamma\gamma$ width of about 0.4 eV \cite{SNA0}\footnote{An alternative approach based on kaon loop leads to a larger value of about $(0.2\sim 0.6)$ keV \cite{ACHASOV}.}. 
These QSSR predictions and the value of the $\sigma\to \gamma\gamma$ direct coupling 
from the data do not favour 
the $\bar qq$ and 4-quark scenarios. However, the conclusion is not sharp as some other approaches may still allow the possibility to have a four-quark state.\\
For further tests of the nature of the $\sigma$, we investigate the extraction of the $\sigma$ coupling to $K^+K^-$. It is clear that, one expects a null value of this coupling in a $\pi\pi$ molecule and/or tetraquark assignements for the $\sigma$. This is in contrast with  its gluonium assignement, where one, instead, 
expects its large (almost) universal coupling to pairs of pseudoscalar mesons \cite{VENEZIA,SNG,SN06} (see Table \ref{tab:S2})\footnote{This typical non-perturbative prediction differs from a perturbative argument (which should not apply below 1 GeV) where the coupling behaves like the current quark mass.}:
\beq
|g_{\sigma K^+K^-}|\simeq |g_{\sigma \pi^+\pi^-}|\simeq 5~{\rm GeV}~.
\label{eq:letcoupling}
\eeq

\vspace*{-0.3cm}
\section{The $g_{\sigma K^+K^-}$ coupling  from  $\pi^+ \pi^-\to \pi\pi/K\bar K$}\vspace*{-0.25cm}
 \nin
Unlike $g_{\sigma \pi^+\pi^-}$, this coupling cannot be directly measured due to phase space suppression.
In the following, we extend the method in \cite{mennessier,MNO} used for elastic $\pi\pi$ scattering to determine the
$\sigma$ parameters presented in the previous section. We shall also use S-matrix coupled-channel models with poles discussed in \cite{KLL}. 
\subsection{The analytic K-matrix model of \cite{mennessier,MNO}}\label{sec:Kmatrix}
\nin
The strong processes are expressed by a K
matrix model representing the $\pi\pi\to\pi\pi/ K\bar K$ amplitudes by a set of resonance poles \cite{mennessier}.
In this case, the dispersion relations in the multi-channel case 
can be solved explicitly, which is not possible otherwise. 
This  model can be reproduced by a set of Feynman diagrams, including 
resonance (bare) couplings to  $\pi\pi$ and $K\bar K$
 and 4-point $\pi\pi$ and $K\bar K$ interaction vertices.  
A subclass of bubble pion
 loop diagrams including resonance poles in the s-channel are resummed
 (unitarized Born).
In \cite{MNO}, we study the elastic $\pi\pi$ scattering, where we introduce 
a {\it shape function} $f_0(s)\equiv f_P(s)$ which  multiplies the $\sigma\pi\pi$ coupling and,
for simplicity, we do not include the 4-point coupling term.  Unlike approaches based on dispersion
relations, this approach can provide a separation of the ``direct" resonance coupling with the ``rescattering" contributions which have been explicitly analyzed for $\gamma\gamma$ scattering in \cite{MNO}\footnote{A separation of the direct and rescattering term can also be studied by measuring $C$ assymetry in $e^+e^-$ \cite{LAYSSAC}.}. In the following, we
discuss this approach (for a pedagocical reason) for the case of 1 channel $\oplus$ 1 resonance. 
The real analytic function $f_P(s):~P\equiv \pi,K$ is regular for  $s > 0$
and has a left cut for  $s \le 0$. For our low energy approach, a convenient
approximation, which allows for a zero at $s=s_{AP}$
and a pole at $\sigma_P>0$ simulating the left hand cut, is: 
\beq
f_P(s)=\frac{s-s_{AP}}{s+\sigma_{D0}} \label{formfactor}~.
\eeq 
The unitary 
$PP$ amplitude is then written as:
\beq
  T_P^{(0)}(s) = \frac{G_P f_P(s)}{s_R-s -  G_P \ftilde_P (s)} = \frac{G_P f_P(s)}{\calD_P(s)}~,
\label{tpipi}
\eeq 
where the index $0$ corresponds to $I=0$,
$T_P^{(0)}=e^{i\delta^{(0)}_P}\sin\delta^{(0)}_P/\rho_P(s)$ with 
 $ \rho_P(s) =({1 - 4 m^2_P/s})^{1/2}$;  $G_P=g_{P,B}^2\equiv g^2_{\sigma PP}/(16\pi)$
 are the bare coupling squared and :
\beq
{\rm {Im}}~ \calD_P = {\rm{Im}} ~ (- G_P \ftilde_P) = - (\therho_P) G_P \ f_P~,  
\label{eq:imaginary}
\eeq
with: $(\theta\rho_P)(s)=0$ below and $(\theta\rho_P)(s)=\rho_P(s)$
above threshold $s=4m_P^2$. 
The amplitude near the pole $s_0$ where $ {\cal D}_P(s_0)=0$ and
$\calD_P(s)\approx \calD'_P(s_0) (s-s_0)$ is:
\beq
  T_P^{(0)}(s)\sim \frac{g_P^2}{s_0-s}; \qquad g_P^2=\frac{G_P
f_P(s_0)}{-\calD'(s_0)}~.
\label{eq:gpi2}  
\eeq
The real part of $\calD_P$ is obtained from a dispersion relation with
subtraction at $s=0$ and one obtains:
\beq
\ftilde_P(s) = \frac{2}{\pi} \Big{[} h_P(s) \ -h_P(0) \Big{]}~:
\label{eq:ftilde}
\eeq
$h_P(s) =f_P(s) \tilde L_{s1}(s)$--$(\sigma_{NP}/(s+\sigma_{DP}))\tilde L_{s1}(-\sigma_{DP})$, 
$\sigma_{NP}$ is the residue of $f_P(s)$ at $-\sigma_{DP}$ and: $\tilde L_{s1}(s) =  
 \big{[}\ga{s - 4 m_P^2}\dr/{m_P^2} \big{]}
\tilde L_1(s,m_P^2)$ with $\tilde L_1$ from \cite{mennessier}. This analysis can be generalized
to the case of 2 channels $\oplus$ 2 resonance poles. A priori, the shape functions differ for the $\pi\pi$ and $K^+K^-$ channels. However, evoking $SU(3)$ symmetry, we can assume (to a first approximation) that they are equal. 
\subsection{Coupled channel model of \cite{KLL}}
 \nin
In the 2- and 3-coupled channel approach presented in \cite{KLL} resonances correspond to 
the closest to physical region poles of the $S$-matrix in complex energy plane.
For each resonance such a pole is chosen among $2^i$ poles ($i$ is number of coupled
channels) what is in contrast with $K$-matrix models and those using Breit-Wigner formulae.
In this way, the unitary $S$-matrix approach in \cite{KLL} delivers us a spectrum of scalar 
mesons below 1.6 GeV together with their couplings and branching ratios fitted to 
experimental data on the $\pi\pi$ and $K\bar K$ phase shifts and inelasticities.
For example, for the Solution A in \cite{KLL}, the ratio of couplings to the $K \bar K$ and $\pi\pi$ channels
for $\sigma$ and $f_0(980)$ states were 0.25 and 2.24 respectively. \\
These fits can be  improved by using the method in \cite{GKPY} based on dispersive
analysis of experimental data.
The authors have shown that theoretical constraints given by Forward Dispersion Relations
(FDR),  sum rules and by once and twice subtracted dispersion
relation (GKPY and the Roy's equations respectively) allow to determine the $\pi\pi$ scattering
amplitudes consistent with analyticity, unitarity and crossing symmetry.
\\
Forward Dispersion Relations (calculated at $t=0$) are used in \cite{GKPY} for three isospin
combinations of $\pi\pi$ amplitudes: for $\pi^0\pi^+$, $\pi^0\pi^0$ and for $t$-channel one
with isospin $I_t=1$.
For example, the FDR for the former two combinations which need two subtractions read:
\vspace*{-0.08cm}
\begin{equation}
    \begin{array}{rl}
&ReF_i(s,0) - F_i(4m_{\pi}^2,0) = \\
&{s(s-4m_{\pi}^2)\over \pi}
-\hspace{-0.5cm}\displaystyle \int \limits_{4m_{\pi}^2}^\infty 
\frac{(2s'-4m_{\pi}^2)ImF_i(s',0)ds'}{s'(s'-s)(s'-4m_{\pi}^2)(s'+s-4m_{\pi}^2)},
\end{array}
\label{FDR12}
\end{equation}
\vspace*{-0.19cm}
where $F_i$ stands for  $\pi^0\pi^+$ or $\pi^0\pi^0$ amplitudes, while
FDR for the latter one do not need subtractions.
In the ideal situation, the difference on the left hand side of Eq. (\ref{FDR12}) 
equals to zero but for realistic 
amplitudes this difference is minimized in the fitting procedure described in \cite{GKPY}.
\\
In the case of nonforward dispersion relation (the Roy's and GKPY equations) derived 
with imposed crossing symmetry condition, a
difference between "output" and "input"
partial wave amplitudes is minimized together with that for FDR.
The "output" ones are calculated for three partial waves $JI$: $S0$, $P1$ and $S2$ up to 
almost 1 GeV and are given by  
\begin{equation}
    \begin{array}{rl}
 &    \mbox{Re } f_{\ell}^{I (output)}(s)  =  ST(a_0^0, a_2^0) \,+
  \\
 &     \displaystyle \sum\limits_{I'}^{}
        \displaystyle \sum\limits_{\ell'}^{}
     \hspace{0.1cm}-\hspace{-0.55cm}
        \displaystyle \int \limits_{4m_{\pi}^2}^{s_{max}}\hspace*{-0.3cm}\ ds'
     K_{\ell \ell^\prime}^{I I^\prime}(s,s') \mbox{Im }f_{\ell'}^{I^\prime}
     (s') + d_{\ell}^{I}(s,s_{max}).
\end{array}
\label{RoyEquations}
\end{equation}
The $a_0^0$ and $a_2^0$ are the $S0$ and $S2$ scattering lengths, 
$K_{ll'}^{II'}(s,s')$ are kernels and $d_l^I(s,s_{max})$ are the so-called driving terms.
The subtracting terms ($ST$) are linear combinations of scattering lengths in the GKPY equations
and do not depend on $s$.
In the Roy's equations, these are also combinations of $a_0^0$ and $a_2^0$ but depend linearly 
on $s-4m_{\pi}^2$.
The integrals with kernels $K_{ll'}^{II'}(s,s')$ are calculated for partial
waves with $l'<2$ for the Roy's equations and with $l'<4$ for GKPY ones.
The maximal value of the effective two pion mass squared $s_{max}$, up to which
experimental phase shifts and inelasticities for waves with $I'$ and $l'$ 
were parameterized, was chosen to be 1.42 GeV.
The driving terms $d_l^I(s,s_{max})$ describe the influence of higher partial waves in the whole $s$
range and of the lower ones above $s = s_{max}$ where Regge model was applied. The ``input" amplitudes are those of which imaginary parts were used
in once and twice subtracted dispersion relations in kernel and driving terms.
\\
As was shown in \cite{GKPYconf} the GKPY equations give output amplitudes with much smaller 
errors than those from the Roy's ones. 
It means that the GKPY equations impose stronger constraints on studied amplitudes.
In addition to the Roy's and GKPY equations, two sum rules which relate high energy (Regge)
parameters to low energy $P$ and $D$ waves were also considered. 
\\  
Following this method, we have performed similar fit for the $S0$ amplitude from 
2- and 3-coupled channel model of \cite{KLL}. 
All other $\pi\pi$ partial waves were the same as in \cite{GKPY} and fixed.
As a result, the ratio of couplings to the $K \bar K$ and $\pi\pi$ channels becomes 0.75
for $\sigma$ and 1.98 for $f_0(980)$, which we report in Table \ref{tab:coupling}.
The change of this ratio for $\sigma$ in comparison with that for solution A of \cite{KLL}
confirms the general conclusion from \cite{GKPY} and \cite{PYI} that available $\pi\pi$ experimental data sets, 
(also this from \cite{klr} used in fits of \cite{KLL}) do not fulfill enough well 
theoretical constraints from dispersion relations.
It shows how important are theoretical demands given by dispersion relations and sum rules.  
Such constraints should be used in the fits of the experimental data of $\pi\pi$ amplitudes.
 \subsection{Data input and results}
 \nin
 We have used the data of inelasticities and phase shifts from \cite{COHEN,ETKIN,HYAMS,KLL}.
 More details of the analysis will be published elsewhere.
In Table \ref{tab:coupling}, we give  the different couplings coming from the analyses
based on the models in \cite{mennessier,MNO} and \cite{KLL}. We shall use the 
normalization in Eq. (\ref{eq:scalarwidth}), implying that $|g_P|^2$ defined in section \ref{sec:Kmatrix} is equal to $|g_{\sigma PP}|^2/(16\pi)$ in the narrow width approximation. We shall also use the relations:
\beq
|g_{S\pi^+\pi^-}|^2={2\over 3} |g_{S\pi\pi}|^2~,~~~~|g_{SK^+K^-}|^2={1\over 2} |g_{SK\bar K}|^2~.
\eeq
We compare in the same Table \ref{tab:coupling}, the results with the ones obtained in \cite{ZHENG09} from the use of dispersion relations in $\pi\pi$ and $K\bar K$ coupled-channel analysis in \cite{ZHENG2}.   We also compare the results with the ones from $\phi\to \sigma/f_0(980)~\gamma$ \cite{BUGG} and $J/\psi
\to\phi~ \pi\pi/ K\bar K$ \cite{BES} decays. One
can notice that the different predictions for the $\sigma$ couplings are quite stable. The $\sigma$ couplings results using the model of \cite{MNO} have been obtained in the case with 1 resonance $\oplus$ 2 channels. In this case, one has obtained two output poles but the 2nd one stays in an unphysical region. The results for the absolute value of the $f_0(980)$ couplings depend on the different models but the ratio of couplings is also quite stable. These results indicate that the $\sigma$ and $f_0(980)$ have important couplings to $K\bar K$, where the one for the $\sigma$ is more remarkable.
\vspace*{-0.3cm}
{\scriptsize
\begin{table}[hbt]
\setlength{\tabcolsep}{0.1pc}
 \caption{\scriptsize    Modulus of the $\pi^+\pi^-$ and $K^+K^-$ complex couplings in GeV of the $\sigma $ and of $f_0(980)$ from S- and K-matrix models for $\pi\pi\to \pi\pi/K\bar K$ scatterings compared with the ones from $\phi$ and $J/\psi$ decays. $r_{S\pi K}\equiv  g_{S K^+K^-}/g_{S\pi^+\pi^-}$: $S\equiv \sigma,~f_0$. $\approx$ means that this result has been obtained in the case 1 resonance $\oplus$ 2 channels.
    }
\begin{tabular}{lccccc}
&\\
\hline
 Processes&$g_{\sigma\pi^+\pi^-}$ & $r_{\sigma\pi K}$ &$g_{f_0\pi^+\pi^-}$ &$r_{f_0\pi K}$&Models  \\
\hline
{ This work} &\\
$\pi\pi\to \pi\pi/K\bar K$&1.55&$\approx 1$&1.57&1.8&\cite{mennessier,MNO} \\
&$2.5$&0.75&$-$&1.98&\cite{KLL} \\
{ Others }&\\
$\pi\pi\to \pi\pi/K\bar K$ &2.5&0.62&1.55&1.2&\cite{ZHENG09}\\
$\phi\to \sigma /f_0(980)~\gamma$&$-$&0.67&$-$&$-$& \cite{BUGG}\\
$J/\psi\to\phi~ \pipi/K\bar K$&$-$&$-$&2.35&1.8& \cite{BES}\\
{ Average}&$2.2$&$0.8$&$1.8$&$1.7$& \\
\hline
\end{tabular}
\label{tab:coupling}
\end{table}
}
\nin

\section{Comparison with the LET results} \label{sec:let}
\subsection{LET for the $\sigma$ couplings to Goldstone bosons}
 \nin
These couplings can be obtained from the vertex function:
\beq
V[q^2\equiv (q_1-q_2)^2]\equiv \la \pi_1\vert \theta_\mu^\mu\vert \pi_2\ra~,
 \eeq
 where:
 \beq
  \theta_\mu^\mu\equiv {1\over 4}\beta(\alpha_s)G^{\mu\nu}_aG_{\mu\nu}^a+\sum_{i\equiv u,d,s}(1+\gamma(\alpha_s))m_i\bar\psi_i\psi_i~,
  \eeq
  is the conformal anomaly (trace of the energy momentum tensor) with $\beta$ and $\gamma_m$ are the QCD $\beta$-function and mass-anomalous dimension while $G^{\mu\nu}$ is the gluon field strength and $\psi_i$ is the quark field. $V(q^2)$
 obeys a once subtracted dispersion relation \cite{VENEZIA}:
 \beq
 V(q^2)=V(0)+{q^2}\int_{4m_\pi^2}^\infty {dt \over t}{1\over (t-q^2-i\epsilon)}~{1\over \pi}{\rm Im} V(t)~,
 \eeq
with the condition:  $V(0)={\cal O}(m^2_\pi) \rar 0$ in the chiral limit. Using also the fact that $V'(0)=1$,  one can then derive the two sum rules:
\beq
\sum_{S=\sigma_B,...} \hspace{-0.3cm} g_{S\pi^+\pi^-}\sqrt{2}f_S=0,~~
{1\over 4} \hspace{-0.1cm}\sum_{S=\sigma_B,...}  \hspace{-0.3cm} g_{S\pi^+\pi^-}{\sqrt{2}f_S\over M_S^2}=1,
\eeq
where $f_S$ is the decay constant analogue to $f_\pi$. The 1st sum rule requires the existence of at least two resonances coupled strongly to $\pi\pi$. Considering the $\sigma_B$ and $\sigma'_B$ (its first radial excitation) but neglecting the small $G$-coupling to $\pi\pi$
as indicated by GAMS \cite{GAMS}\footnote{The $G(1.6)$ is expected
to  couple mainly to the $U(1)_A$ channels $\eta'\eta'$ and through mixing to $\eta\eta',~\eta\eta$ 
\cite{VENEZIA,SNG}.} and some other data \cite{PDG}, one predicts in the chiral limit \cite{VENEZIA,SNG} 
the almost universal couplings in Eq. (\ref{eq:letcoupling}) (see Table \ref{tab:S2}), which would correspond to a large width~\footnote {Notice that the analysis in Ref. \cite{SNG} also indicates  that $\sigma_B$ (in the real axis) having a mass below 750 MeV cannot be wide ($\leq 200$ MeV) due to the sensitivity of the coupling to $M_\sigma$. }:
\beq
\Gamma_{\sigma_B\to\pi^+\pi^-} \simeq 0.5~ {\rm GeV}~.
  \label{eq:sigmawidth} 
\eeq
This large width into $\pi\pi$ is a typical OZI-violation  due to non-perturbative effects expected to be important in the region below 1 GeV, where perturbative arguments (like a vanishing of the hadronic glueball coupling in the chiral limit \cite{CHANOW}) are valid in the region of the G(1.5-1.6) 
cannot be applied. 
  \vspace*{-0.1cm}
\subsection{Evidences that the $\sigma$ is a gluonium} 
 \nin
In section \ref{sec:qssr}, we have given several indications that the $\sigma$ can be a gluonium
from the values of its  small $\gamma\gamma$  and large hadronic widths and of its on-shell mass.
In the previous section, the value of its $K^+K^-$ coupling, which is about 0.7 times of its $\pi^+\pi^-$  
one (see Table \ref{tab:coupling}), gives a further support on its gluonium structure due to the fact that
a $\bar \pi\pi$ molecule and a tetraquark assignements would lead to a null value of the $K^+K^-$ coupling. This gluonium nature of the $\sigma$ can indicate that the effective $\bar KK \bar\pi\pi$ four-vertex used in a kaon loop model for successfully explaining e.g. $\phi\to\gamma\pi\pi$ can be induced by a s-channel exchange of a gluonium state.  Indeed, using a gluonium picture, one can predict correctly this observed radiative width \cite{SN09,SNG,SN06}.
For better comparing the results obtained in the complex plane with the QSSR results obtained in the real axis, we introduce like in \cite{MNO} the on-shell meson masses and
hadronic widths \cite{SIRLIN}, where the amplitude is purely imaginary 
at the phase 90$^0$: 
\beq
{\rm Re} {\cal D}({(M^{\rm os}_\sigma})^2)=0\lrar M_\sigma^{\rm os}\approx 0.92~{\rm GeV}~. 
\eeq
In the same way as for the mass, one can define an ``on-shell width"
 [see Eqs. (\ref{eq:imaginary}) and (\ref{eq:gpi2})] evaluated at $s=(M^{\rm os}_\sigma)^2$ :
\beq
M^{\rm os}_\sigma \Gamma^{\rm os}_\sigma\simeq {{\rm Im}~ \calD\over -{\rm Re~} {\cal D}'}
\lrar \Gamma_{\sigma\to\pi^+\pi^-}^{\rm os}\approx 0.7~{\rm GeV}~,
\eeq
which are comparable with the Breit-Wigner mass and width \cite{HYAMS,ESTABROOKS,AMP}:
\beq 
M_{BW}\approx \Gamma_{BW}\approx 1~{\rm GeV}.
 \label{sigmamass} 
\eeq 
These values lead to the on-shell coupling:
\beq
\vert g_{\sigma\pi^+\pi^-}^{\rm os}\vert \simeq 6~{\rm GeV}~,~r_{\sigma\pi K}\equiv 
{g_{\sigma K^+K^-}^{\rm os}\over g_{\sigma\pi^+\pi^-}^{\rm os}}\simeq 0.8~.
\label{sigmacoupl}
\eeq
These fitted values and the one predicted in Eq. (\ref{eq:letcoupling}) strongly indicate a large gluonium 
component in the $\sigma$ wave function. 
\vspace*{-.15cm}
\section{Summary and conclusions}
\vspace*{-0.25cm}
 \nin
\b We have extracted the $\sigma\pi^+\pi^-$ and $\sigma K^+K^-$ couplings from $\pi\pi\to \pi\pi/K\bar K$ scatterings using different models of K- and S-matrices. We have shown in Table \ref{tab:coupling} the 
different results and have compared them with the ones from $\phi\to \sigma/f_0(980)~\gamma$ and $J/\psi
\to \phi~ \pi\pi/K\bar K$ decays. A comparison of these
results with the predictions from LET$\oplus$QSSR favours a large gluonium/glueball content in the $\sigma$ wave function. \\
\b This feature can explain the splitting of its mass from the pion one, which is similar to the $\pi-\eta'$ mass splitting occuring in the well-known $U(1)_A$ anomaly \cite{WITTEN}. The similarity with the $\eta'$ is also signaled by the affinity of the $\sigma$ to couple to Goldstone bosons indicating a large OZI violation in its decay into $\pi\pi$ and a large value of its $K\bar K$ coupling. In this way, the $\sigma$ associated to the trace of the energy-momentum tensor $\theta^\mu_\mu$ [$U(1)_V$ conformal anomaly] can be considered as the partner of the $\eta'$ of the $U(1)_A$ anomaly. This non-$\bar qq$ property of the $\sigma$ may also go in lines with the $1/N_c$ counting done in \cite{PELAEZ} using unitarized ChPT partial waves for describing $\pi\pi$ scatterings. \\ 
\b Therefore, we consider that the assignement for the $\sigma$ as a non-strange partner of the $\kappa/K^*_0(800)$ (if confirmed), like often advocated in the four-quark literature (see e.g. \cite{SCHECHTER,MONTANET}), may not be appropriate. The latter having an isospin 1/2 cannot have a gluonium in its wave function.  Indeed, the $\kappa$ can likely be the isoscalar partner of the $a_0(980)$, where, in the standard classification of current algebra, they are respectively the $\bar us$ and $\bar ud$ mesons having a mass around 1 GeV \cite{SN06,SNG} which are associated to the divergence of the vector currents. Also, within this description, a successful determination of the running light quark mass differences from QSSR have been achieved \cite{SNB,SNQ,SCAL}.\\
\b Finally, with the gluonium nature of the $\sigma$ and its large coupling to pseudoscalar boson pairs, the effective four-meson vertex $ K\bar K \pi\pi$ used e.g. for explaining the $\phi\to\gamma\pi\pi$ process can be due to a $s$-channel exchange of a glue rich state. We plan to come back to this point as well as to the study of the gluonium-quarkonium mixing below 1 GeV which has been initiated in \cite{BN,SNG}.

\vspace*{-0.35cm}
\section*{Acknowledgements} 
\vspace*{-0.25cm}
\nin
One of us (R.K.) would like to thank S.N. for the kind hospitality at the CNRS and at the University of Montpellier.  
S.N. has been partially  supported by the CNRS-IN2P3 within the French-China Particle Physics Laboratory (FCPPL). He thanks  X-Q. Li, Q. Zhao and H-Q. Zheng for  the hospitality at 
the University of Nankai (Tianjin), at IHEP (Beijing) and at Peiking University  (Beijing). We thank D. Bugg, W. Ochs,  P. Minkowski, X-G. Wang and H-Q. Zheng for some communications. This work has been initiated at the QCD 08 Montpellier Conference (July 2008).

\vspace*{-0.25cm}
\input{bib_sigmab}

\end{document}